\documentclass[twocolumn,amsmath,amssymb,10pt,aps,floatfix,superscriptaddress]{revtex4}

\usepackage{graphicx}
\usepackage{dcolumn}
\usepackage{bm}
\usepackage{graphicx}  
\usepackage{dcolumn}   
\usepackage{bm}        
\usepackage{verbatim}   
\usepackage[usenames]{color}  
\usepackage{stmaryrd}
\usepackage{float}
\usepackage{units}
\usepackage{textcomp}
\usepackage{textcomp}
\usepackage{amsmath}
\usepackage{commath}
\usepackage{amssymb}
\usepackage{esint}
\usepackage{ae,aecompl}
\usepackage[colorlinks=true,linkcolor=blue,citecolor=blue]{hyperref}
\usepackage{color}

\allowdisplaybreaks

\begin{document}

\title{Seeking Maxwell's Demon in a non-reciprocal quantum ring}
\author{Aram Manaselyan \footnote{amanasel@ysu.am}}
\affiliation{ Department of Solid State Physics, Yerevan State
University, 0025 Yerevan, Armenia}
\author{Wenchen Luo \footnote{luo.wenchen@csu.edu.cn} }
\affiliation{ School of Physics and Electronics,
Central South University, Changsha, Hunan, P.R. China 410083}
\author{Daniel Braak \footnote{d.braak@fkf.mpg.de} }
\affiliation{ Max-Planck Institut f\"ur Festk\"orperforschung,
Heisenbergstra{\ss}e 1, 70569 Stuttgart, Germany}
\author{Tapash Chakraborty \footnote{Tapash.Chakraborty@umanitoba.ca} }
\affiliation{ Department of Physics and Astronomy, University
of Manitoba, Winnipeg, Canada R3T 2N2}

\begin{abstract}
A non-reciprocal quantum ring, where one arm of the ring contains the Rashba
spin-orbit interaction but not in the other arm, is found to posses very unique
electronic properties. In this ring the Aharonov-Bohm oscillations are totally
absent. That is because in a magnetic field the electron stays in the
non-Rashba arm, while it resides in the Rashba arm for zero (or negative)
magnetic field. The average kinetic energy in the two arms of the ring are
found to be very different. It also reveals different ``spin temperature" in
the two arms of the non-reciprocal ring. The electrons are sorted according
to their spins in different regions of the ring by switching on and off (or
reverse) the magnetic field, thereby creating order without doing work on
the system. This resembles the action of a demon in the spirit of Maxwell's
original proposal, exploiting a non-classical internal degree of freedom.
Our demon clearly demonstrates some of the required features on the nanoscale.
\end{abstract}

\maketitle

In the middle of nineteenth century, physicists were grappling with the
implications of the newly established second law of thermodynamics, whose one
profound proclamation was that, it is not possible for heat to spontaneously
flow from a cold system to a hot system without external work being done on the
system (Clausius statement). Then in 1867, James Clerk Maxwell questioned
\cite{max_book} if the above statement was true only for a system whose
properties were governed by the average behavior of the particles or whether it
is still valid even at the level of individual particles. In order to test this
idea, Maxwell conceived the famous {\it gedankenexperiment} where a tiny
intelligent being ({\it demon}) with exceptional capacity of being able to
observe individual molecules and their speed, was assigned a very special job.
The demon was to sort, in a closed box of uniformly distributed particles, the
fast (hot) particles and slow (cold) particles in two compartments of the box
separated by a wall with a tiny trap door. As time passes, the demon opens or
shuts the door to allow the fast or slow particles in their respective
compartments, thereby creating a temperature gradient from the initial uniform
temperature without doing work on the system and thus violate the second law
of thermodynamics \cite{Lieb,capek}.

Given the paucity of intelligent demons, for 150 years Maxwell's demon
remained an interesting idea that could never be implemented in a real world.
However, in recent years, rapid progress in nanoscale physics has resulted in
devices where the dynamics of even individual electron can be controlled.
Examples of such systems are quantum dots (artificial atoms)
\cite{maksym,dot_book} and quantum rings \cite{chak_ring_1,chak_ring_2,%
chak_ring_3,ring_book}. Quite naturally, this demon with incredible dexterity
has now made a remarkable comeback in nanoscale devices. Theoretical studies
indicate that the demon is lurking in quantum dots (QDs) \cite{QDD}, quantum
Hall systems (`chiral demon') \cite{gloria}, and even in a photonic setup
\cite{photonic}. Experimentally, Pekola et al. \cite{pekola} have reported
creating Maxwell's demon in a system of two quantum dots.

\begin{figure}
\includegraphics[width=4cm]{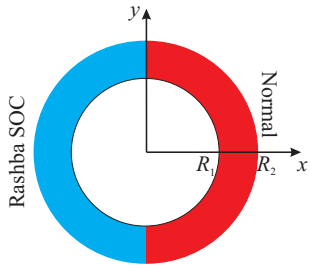}
\caption{A non-reciprocal quantum ring where one arm contains the Rashba
spin-orbit coupling (SOC) while that coupling is absent in the other arm.}
\label{man_ring}
\end{figure}

Our nanoscale system, a {\it non-reciprocal} quantum ring (Fig.~\ref{man_ring})
\cite{mannhart_1,mannhart_2}, where one arm of the ring contains the Rashba
spin-orbit coupling (RSOC) \cite{rashba_1,rashba_2,rashba_3,rashba_ring} while
the other arm is normal, i.e., without the RSOC, is shown below to behave as
if the demon is at work in pursuit of its assignment to break ``one of the most
perfect laws in physics" \cite{Lieb}. In what follows, we demonstrate that
electron can be channeled through the two arms according to their spin states
by switching on and off (or reverse) an external magnetic field. Most
significantly, this leads to different kinetic energies and effective spin
temperature in the two arms in the ground state, not unlike the expected
signature of a Maxwell Demon. The demon neither infuses energy into the system
or create entropy on its own, nor does it process information. In this sense
it is quite different from the usual proposals \cite{QDD,gloria,photonic,%
pekola} based on a variant of Landauer's principle \cite{landauer}.
Nevertheless it acts effectively as a one-way trapdoor between two spatially
separated regions of the system.

\section{Results}

Let us consider a non-reciprocal two-dimensional quantum ring (QR) with inner
radius $R^{}_1$ and outer radius $R^{}_2$ and with the Rashba spin-orbit
interaction \cite{rashba_1,rashba_2,rashba_3,rashba_ring} that is applied only in
one half of the ring. For simplicity we choose the confinement potential of the QR
with infinitely high borders: $V^{}_\mathrm{conf}(r)=0$, if $R^{}_1\leq r
\leq R^{}_2$ and infinity otherwise.

\paragraph{Model}
The single-particle Hamiltonian of the system in an external magnetic field
applied in the growth direction (the $z$ direction, perpendicular to the ring)
is
\begin{equation}\label{singleH}
    {\cal H}= {\frac1{2m^{}_e}}\mathbf\Pi^2+
V^{}_\mathrm{conf}(r)+ \frac12 g\mu^{}_\mathrm{B} B\sigma^{}_z+ {\cal H}^{}_1,
\end{equation}
where $\mathbf\Pi=\mathbf p+\frac ec \mathbf A$, $\mathbf{A}=B/2\left(-y,x,0
\right)$ is the vector potential of the applied magnetic field along the $z$
axis in the symmetric gauge. The third term on the right hand side of
(\ref{singleH}) is the Zeeman splitting. The last term describes the
non-reciprocal RSOC
\begin{equation}\label{NR}
    {\cal H}^{}_1 = f(\theta)H^{}_\mathrm{SO} f(\theta)
\end{equation}
where $f(\theta)=1$ if $\pi/2\leq \theta \leq 3\pi/2$ and 0 otherwise, is a
function which describes the region of the QR where the RSOC is applied.
$H^{}_\mathrm{SO}$ is the RSOC Hamiltonian \cite{rashba_1,rashba_2,rashba_3,%
rashba_ring,Winkler} \begin{equation}\label{Rashba}
H^{}_\mathrm{SO}=\frac \alpha\hbar \left[\bm{\sigma}\times\mathbf{\Pi}
\right]^{}_z=i \frac{\alpha}{\hbar}\left(\begin{array}{cc}
    0 & \Pi^{}_- \\
    -\Pi^{}_+ & 0
\end{array}\right),
\end{equation}
with $\Pi^{}_{\pm}=\Pi^{}_x \pm i\Pi^{}_y$ and $\alpha$ being the RSOC
parameter which depends on the asymmetry in the $z$ direction, generated
either by the confinement or the electric field. The SOC coupling strength
$\alpha$ can be experimentally determined for the QR materials
\cite{rashba_1,rashba_2,rashba_3}. Practical realization of the Rashba
arm of the ring could perhaps be possible by appropriate covering of the
ring by the gate electrode \cite{nitta_apl}.

We employ the exact diagonalization scheme \cite{exactd} in order to find the
eigenvalues and eigenfunctions of the Hamiltonian (\ref{singleH}). We take as
basis states the eigenfunctions of the Hamiltonian (\ref{singleH}) at $B=0$ and
$\alpha=0$ \cite{ring_book,majorana,zno}. The eigenfunctions of this Hamiltonian
then have the form

\begin{equation}
\phi^{}_{nl}(r,\theta)=\frac{C}{\sqrt{2\pi}}e^{\mathrm{i}l\theta}\left(J^{}_l
(\gamma^{}_{nl}r)-\frac{J^{}_l (\gamma^{}_{nl}R^{}_1)}{Y^{}_l(\gamma^{}_{nl}
R^{}_1)}Y^{}_l(\gamma^{}_{nl}r)\right)\\
\label{Basis}
\end{equation}
where $n=1,2,\ldots$, $l=0,\pm1,\pm2,\ldots$ are quantum numbers, $J^{}_l(r)$
and $Y^{}_l(r)$ are Bessel functions of the first and second kind respectively,
$\gamma^{}_{nl}=\sqrt{2m^{}_eE^{}_{nl}/\hbar^2}$, where $E^{}_{nl}$ are the
eigenstates defined from the boundary condition at $r=R^{}_2$, and the
constant $C$ is determined from the normalization integral.

\begin{figure}
\includegraphics[width=4cm]{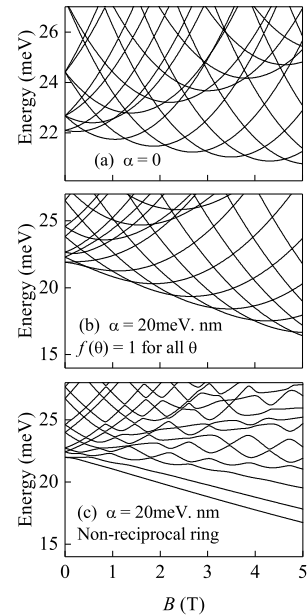}
\caption{Magnetic field dependence of the electron energy spectra (a) for the
QR without the RSOC, (b) for normal QR with the RSOC parameter $\alpha= 20$
meV.nm and (c) for non-reciprocal QR with RSOC parameter $\alpha=20 $meV.nm.}
\label{energy}
\end{figure}

In order to evaluate the energy spectrum and the wave functions of the
Hamiltonian (\ref{singleH}) we need to digonalize the matrix of the Hamiltonian
(\ref{singleH}) in a basis (\ref{Basis}). Numerical calculations are performed
for InAs QR with parameters \cite{Winkler} $m^{}_e=0.042m^{}_0$ ($m^{}_0$ is the free electron
mass), $g= -14$, $R^{}_1=300$\AA\ and $R^{}_2=500$\AA. In our calculations we
have used 210 single-electron basis states $\left|n,l,s\right>$ where
$s=\pm1/2$ is the electron spin, which is adequate for determining the first
few energy eigenvalues of our system with high accuracy.

\paragraph{Numerical results}
In Fig.~\ref{energy} the energy spectra of the system against the magnetic
field is presented for various parameters values. For a QR without the RSOC
[$\alpha=0$, Fig.~\ref{energy} (a)] the usual Aharonov-Bohm (AB) oscillations
\cite{ring_book} can be observed. With an increase of the magnetic field, the
ground state periodically changes, as expected. For the normal QR with the RSOC
[$f(\theta)=1$ for all $\theta$, see Fig.~\ref{energy} (b)] the AB oscillations
are still present, albeit with smaller amplitude but with the same period.
The degeneracy of the energy levels at $B=0$ being partially lifted, while
the appearing level crossings as function of $B$ are due to rotational
invariance. This invariance is broken in the non-reciprocal ring and all
degeneracies are lifted, leading to the complete disappearance of the AB
oscillations in the ground state [Fig.~\ref{energy} (c)]. Clearly this
means that the electron confinement inside the non-reciprocal ring is very
different from that of the other two cases.

\begin{figure}
\includegraphics[width=7cm]{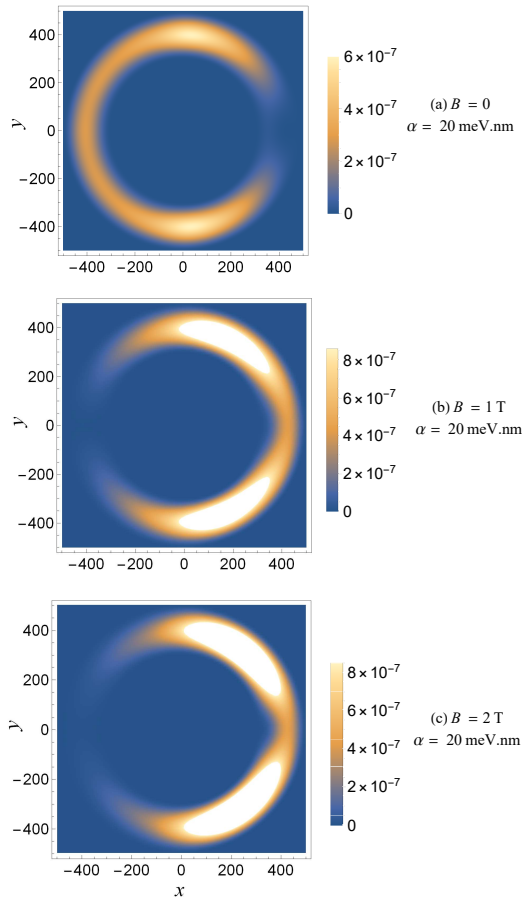}
\caption{Ground state electron density in a QR with non-reciprocal RSOC for
various values of the magnetic field.}
\label{density_1}
\end{figure}

In Fig.~\ref{density_1} the ground state electron density in a QR with
non-reciprocal RSOC is presented for various values of the magnetic field. In
the absence of the magnetic field [Fig.~\ref{density_1} (a)] the electron
charge is distributed all over the ring with a slight shift into the arm of
the ring where the RSOC is present. However, when the magnetic field is
increased, the electron moves to the areas of the ring where the RSOC is
absent [Fig.~\ref{density_1} (b) and (c)]. That is why we cannot observe the
ground state AB oscillations in Fig.~\ref{energy} (c). The reason for this
intriguing phenomena is that in a non-reciprocal QR the arm of the ring with
RSOC is attractive for the spin-down states and repulsive for the spin-up
states, as deduced from the average spin of the ground state $\langle S^{}_z
\rangle$. Without the magnetic field the average spin of the ground
state is almost zero because the RSOC mixes the spin-up and spin-down states.
With increase of the magnetic field the average spin increases and is positive
due to the Zeeman effect. Therefore the electron is confined in the non-Rashba
arm of the ring.

\begin{figure}
\includegraphics[width=6cm]{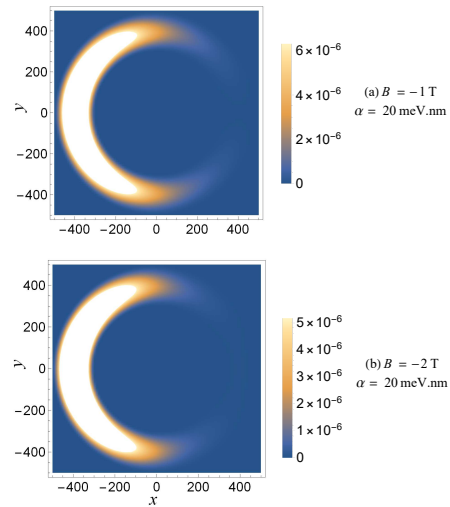}
\caption{Same as in Fig.~\ref{density_1}, but for opposite direction of
the magnetic field.}
\label{negativeB}
\end{figure}

Figure~\ref{negativeB} is the same as Fig.~\ref{density_1} but for opposite
direction of the magnetic field (negative values of $B$). In this case, when
the magnetic field is increased, the average spin of the electron ground state
is negative and now the electron is confined in the Rashba arm of the ring.
Clearly, the magnetic field acts as the `trap door' for the demon to sort the
electron according to their spins. The magnetic field can be used to control
the electron spin of the ground state and thus the confinement of the
electron in the ring.

\begin{figure}
\includegraphics[width=7cm]{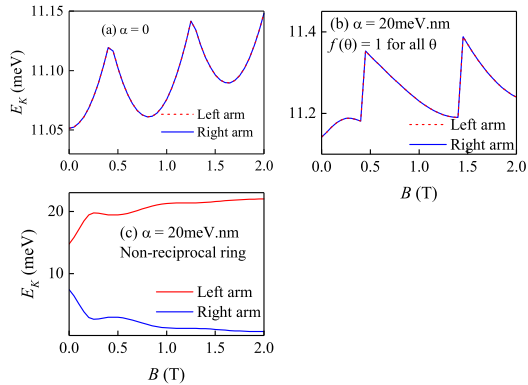}
\caption{Expectation values of the kinetic energy operator in the two arms
of the (a) Rashba free, (b) Rashba and (c) non-reciprocal ring as a function
of the magnetic field.}
\label{kinetic}
\end{figure}

\paragraph{Spin temperatures and the ``Maxwell's demon''}
Let us recall that Maxwell inferred the temperature difference in the two
chambers of his gedankenexperiment from different expectation values
of the kinetic energy (fast molecules in one chamber and slow molecules in the
other chamber) using the standard Maxwell-Boltzmann distribution of a
classical ideal gas. We shall now evaluate the corresponding quantity in our
quantum system. The average kinetic energies in our system for the left and
right arms of the ring can be defined as expectation values of the kinetic
energy operator
\begin{equation}
    E^L_K = \int_{R^{}_1}^{R^{}_2}\int_{\pi/2}^{3\pi/2}\psi^*(r,\theta)
\frac{\Pi^2}{2m^{}_e}\psi(r,\theta) r dr d\theta
\end{equation}
\begin{equation}
    E^R_K = \int_{R^{}_1}^{R^{}_2}\int_{-\pi/2}^{\pi/2}\psi^*(r,\theta)
\frac{\Pi^2}{2m^{}_e}\psi(r,\theta) r dr d\theta
\end{equation}
where $\psi(r,\theta)$ is the ground state wave function of the system. The
obtained average kinetic energies are shown in Fig.~\ref{kinetic}. The
kinetic energies in the two arms of the normal and the Rashba ring are equal,
but differ greatly for all $B$ in case of the non-reciprocal ring. The reason
for this behavior (not foreseeable by Maxwell) is the presence of a quantum
degree of freedom (spin) which is used as a marker by the Hamiltonian itself
(the ``demon" is played by the magnetic field) to separate the particles with
high and low kinetic energy into spatially distinct regions. It is therefore
the simplest manifestation of a demon-like process without the interference of
an intelligent being.The magnetic field could be turned on or reversed
adiabatically, so that no entropy is generated during this process.

\begin{figure}
\includegraphics[width=7cm]{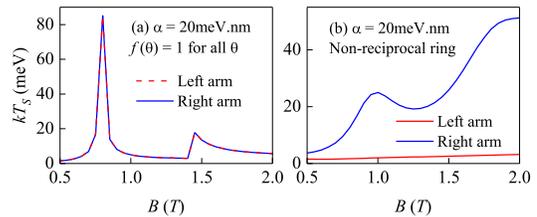}
\caption{Spin temperatures in the two arms of the (a) Rashba and (b)
non-reciprocal ring versus the magnetic field.}\label{Temp}
\end{figure}

Similarly, we can also evaluate the average spin polarizations $\langle S^L_z
\rangle$ and $\langle S^R_z\rangle$ for the left and right arms of the ring as
expectation values of the spin operator for the ground state. To assess the
action of our demon on the spin degree of freedom, we can define ``spin
temperature" $T^{}_S$, using only the Zeeman term in the spin Hamiltonian
\begin{equation}\label{Sz}
    \langle S^{}_z\rangle=\frac12\tanh{\left(\frac{g\mu^{}_B B}{2kT^{}_S}
    \right)}.
\end{equation}
where $k$ is the Boltzmann constant. Eq.~(\ref{Sz}) is used here to define
a canonical ``temperature" associated only with the spin degree of freedom for
arms of the ring, using the ground state values of the spin polarization.
This yields two different spin temperatures in the two arms of the non-%
reciprocal ring (Fig.~\ref{Temp}). Although these temperatures cannot be
associated with the total system, which remains always at $T=0$, the imbalance
between the right and left arm can be expressed as different temperatures of a
``spin gas" in the spirit of Maxwell \cite{max_book}.

\begin{figure}
\includegraphics[width=5cm]{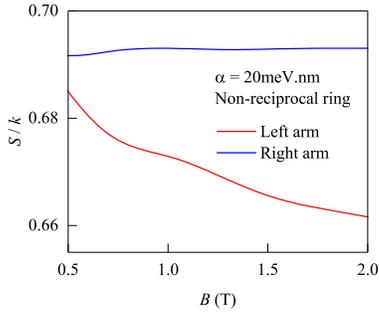}
\caption{Spin entropy in the two arms of non-reciprocal ring versus the magnetic field.}\label{entropy}
\end{figure}

Once the spin temperatures of the different arms are obtained  we can evaluate  the ``spin entropy'' of the two arms \cite{pathria}

\begin{equation}
S_{spin}=k\left\{ \ln \left[ 2\cosh \left( \frac{g\mu ^{}_B B}
{2kT^{}_S}\right) \right] - \frac{g\mu^{} _B B}
{2kT^{}_S}\tanh \left( \frac{g\mu ^{}_B B}{2kT^{}_S}
\right)\right\}.
\end{equation}
It characterizes the demon's action on the internal spin 
of the system. In Fig.~ \ref{entropy}, we notice a sharp drop
in entropy of the ``cool" arm (with the SOC) and a flat
curve in the ``hot" arm (without the SOC). Both temperature and entropy of the spin subsystem indicate the sorting action of the ring but not a deviation from equilibrium of the total system. It is well-known that the density matrix of a subsystem $A$ is usually mixed ($S_{A} > 0$) while the total system remains in a pure state $S_{tot}=0$. In our case, the spin degree of freedom acquires a non-zero entropy because it is entangled with the spatial degrees of freedom through the Rashba-coupling. The crucial difference between the inhomogeneous Rashba-ring and an inhomogeneous potential (which would also lead to a spatially varying electron density) is the fact that the potential $V_{conf}(r)$ is constant along the angle $\theta$. The spin, acting as a quantum marker, is ``measured" by the magnetic field without recording of information and causes the sorting in both halves of the ring.  

\paragraph{Conclusions and Discussions}
In a non-reciprocal QR, the AB oscillations disappear for all
values of the magnetic field. The electron density indicates that, for zero
magnetic field the electron is mostly located in the Rashba arm of the ring.
However, in a non-zero magnetic field the electron moves to the non-Rashba
arm of the ring. The Rashba arm is attractive for states with negative average
spin polarization and repulsive for states where the average spin polarization
is positive. Of course all of our states are mixtures of spin-up and spin-down
states due to the SOC. The ground state electron density indicates that, with
an increase of the magnetic field the value of the average spin for the ground
state is positive and increases. Hence the electron is confined in the
non-Rashba arm of the ring. For negative values of the magnetic field the situation is reversed.

The non-reciprocal QR is clearly a manifestation of demon's workplace: the
electron sorting in different arms of the ring according to their spins leads
to different kinetic energies and spin temperatures in the two arms of the
ring. This unusual sorting by the demon takes place through the spin, which
is a pure quantum character and does not have any classical counterpart.
Observation of the absence of AB oscillations in a non-reciprocal QR
for all values of the magnetic field, and the different temperatures in the
two arms would strongly indicate the {\it action}, if not necessarily the
{\it presence} of the ``restless and lovable poltergeist" \cite{landsberg} in
our nanoscale system. The Coulomb interaction might play an important role in
our ring containing multiple electrons, although AB oscillations are a
single-particle effect.

In the experiment involving two quantum dots
\cite{pekola} mentioned above, the system-demon interaction was mediated by
the Coulomb potential. Our present approach is well suited to study also
the case of interacting electrons very accurately in a non-reciprocal QR.
We have confined our analysis to ground state properties. Obviously, a
generalization to finite temperatures and especially to real-time dynamics
is necessary to investigate possible consequences for the second law of
thermodynamics, which has not been impacted by our results. However,
the demon-like sorting action appears already at $T=0$ in our present system
and might turn out to be useful in algorithmic cooling \cite{algo,aspuru} and
`spin refrigeration' \cite{boykin} in quantum information science.

Although we have considered InAs quantum ring as our model system, for
experimental realization of non-reciprocal quantum rings the material of
choice could also be graphene \cite{boehm_1,boehm_2,graphene_book,%
graphene_review,ssc,xuefeng,abergel,julia}. Graphene quantum rings have been
studied theoretically \cite{abergel_ring,Recher76} and experimentally
\cite{russo}, where observation of AB oscillations was reported. The Rashba
SOC in graphene is negligibly small \cite{gmitra}. However, there are several
proposals in the literature about enhancement of Rashba SOC in graphene
\cite{marchenko,rossi}, that could perhaps help to create the Rashba arm
of the ring.

\section{Methods}
{\small
We use the basis in Eq. (4) to apply the exact diagonalization scheme for the Hamiltonian (1). We use 210 single-electron basis states and 
numerically evaluate all the corresponding matrix elements of the Hamiltonian (1). The eigenvalues and eigenfunctions of the 210x210 
matrix have been evaluated using the standard numerical diagonalization algorithms [23 - 25]. The exact diagonalization scheme has been widely used by many other authors. In this method the energy eigenvalues of complex quantum structures are evaluated with desired 
accuracy by appropriately increasing the number of basis states. This method can therefore be considered as exact (numerically). In our 
present  work, the number of basis states chosen  is sufficient to evaluate the eigenvalues with very high accuracy. The ground state wave functions, the kinetic energies, the spin temperature, and the spin entropy are obtained by numerically evaluating the integrals.}

\vspace*{0.2cm}

{\bf Acknowledgements}\\
{\small
T.C. thanks Jochen Maanhart for helpful discussions. He also thanks Jesko
Sirker for several demon-related ideas. The work of A.M. has been supported
by the Armenian State Committee of Science (Project No. 18T-1C223). W.L.
acknowledges support by the NSF-China under Grant No. 11804396. D.B. thanks
Raymond Fresard for the warm hospitality during his stay at CRISMAT, ENSICAEN,
which was supported by the French Agence Nationale de la Recherche, Grant No.
ANR-10-LABX-09-01 and the Centre national de la recherche scientifique
through Labex EMC3. }

\vspace*{0.2cm}

{\bf Author contributions} \\
{\small
T.C. supervised the project; A.M. wrote the code and performed the numerical
work; T.C. and D.B. aided in the analysis of the results; T.C., A.M., W.L.
and D.B. contributed in designing the project; T.C., A.M., D.B. and W.L.
wrote the manuscript; all authors provided feedbacks on the manuscript. }

{\bf Corresponding author} \\
Correspondence to Tapash Chakraborty (Tapash.Chakraborty@umanitoba.ca) or Wenchen Luo (luo.wenchen@csu.edu.cn).
\vspace*{0.2cm}

{\bf Additional information} \\
{\small
{\bf Competing interests:} The authors declare no competing interests.
}


\newpage
\begin{thebibliography}{9999}

%

\bibitem{max_book}
J.C. Maxwell, {\it Theory of Heat}, (Longman, London, 1871).

\bibitem{Lieb}
E.H. Lieb and J. Yngvason, Phys. Rep. {\bf 310}, 1 (1999).

\bibitem{capek}
V. Capek and D.P. Sheehan, {\it Challenges to the Second Law of 
Thermodynamics, Theory and Experiment}, Springer (2005).

\bibitem{maksym}
P.A. Maksym and T. Chakraborty, Phys. Rev. Lett. {\bf 65}, 108 (1990).

\bibitem{dot_book}
T. Chakraborty, {\it Quantum Dots}, (Elsevier, New York, 2001).

\bibitem{chak_ring_1}
T. Chakraborty, Adv. Solid State Phys. {\bf 43}, 79 (2003).

\bibitem{chak_ring_2}
T. Chakraborty and P. Pietil\"ainen, Phys. Rev. B {\bf 50}, 8460 (1994);

\bibitem{chak_ring_3}
T. Chakraborty and P. Pietil\"ainen, in {\it Transport Phenomena in
Mesoscopic Systems}, edited by H. Fukuyama and T. Ando
(Springer-Verlag, Heidelberg, 1992).

\bibitem{ring_book}
T. Chakraborty, A. Manaselyan and M.G. Barseghyan, in {\it Physics 
of Quantum Rings}, edited by V.M. Fomin (Springer, New York, 2018)
Ch. 11.

\bibitem{QDD}
P. Strasberg, G. Schaller, T. Brandes, and M. Esposito, Phys. Rev. Lett.
{\bf 110}, 040601 (2013).

\bibitem{gloria}
G. Rossello, R. Lopez, and G. Platero, Phys. Rev. B {\bf 96}, 075305 (2017).

\bibitem{photonic}
M.D. Vidrighin, O. Dahlsten, M. Barbieri, M.S. Kim, V. Verdal, and 
I.A. Walmsley, Phys. Rev. Lett. {\bf 116}, 050401 (2016).

\bibitem{pekola}
J.V. Koski, A. Kutvonen, I.M. Khaymovich, T. Ala-Nissila, and J.P. Pekola,
Phys. Rev. Lett. {\bf 115}, 260602 (2015).

\bibitem{mannhart_1}
J. Mannhart, J. Supercond. Novel Magn. {\bf 31}, 1649 (2018).

\bibitem{mannhart_2}
J. Mannhart, P. Bredol, and D. Braak, Physica E {\bf 109}, 198 (2019).

\bibitem{rashba_1}
Y. A. Bychkov and E. I. Rashba, J. Phys. C 17, 6039 (1984).

\bibitem{rashba_2}
D. Grundler, Phys. Rev. Lett. {\bf 84}, 6074 (2000). 

\bibitem{rashba_3}
J. Nitta, T. Akazaki, H. Takayanagi, and T. Enoki, Phys. Rev. Lett. {\bf 78}, 
1335 (1997).

\bibitem{rashba_ring}
H.-Y. Chen, P. Pietil\"ainen, and T. Chakraborty, Phys. Rev. B {\bf 78}, 
073407 (2008).

\bibitem{Winkler}
R. Winkler, Spin-Orbit Coupling Effects in Two Dimensional Electron and Hole Systems (Springer, Berlin, 2003).

\bibitem{landauer}
R. Landauer, IBM J. Res. Dev. {\bf 5}, 183 (1961).

\bibitem{nitta_apl}
J. Nitta, F.E. Meijer, and H. Takayanagi, Appl. Phys. Lett. {\bf 75}, 695 (1999).

\bibitem{exactd}
P. Pietil\"ainen and T. Chakraborty, Phys. Rev. B {\bf 73}, 155315 (2006).

\bibitem{majorana}
A. Ghazaryan, A. Manaselyan, and T. Chakraborty, Phys. Rev. B {\bf 93},
245108 (2016).

\bibitem{zno}
T. Chakraborty, A. Manaselyan and M. Barseghyan, J. Phys.: Condens. Matter
{\bf 29}, 075605 (2017).

\bibitem{pathria}
R.K. Pathria, {\it Statistical Mechanics} (Batterworth Heinemann 1996).

\bibitem{landsberg}
P.T. Landsberg, {\it The Enigma of Time} (Adam Hilger Ltd., Bristol 1982).

\bibitem{algo}
{\it Elements of Quantum Information}, edited by W.P. Schleich and H. Walther
(Wiley-VCH Verlag, Weinheim 2007); {\it Quantum Information and Computation 
for Chemistry}, edited by S. Kais (John Wiley \& Sons, 2014).

\bibitem{aspuru}
J.-S. Xu, et al., Nat. Photonics {\bf 8}, 113 (2014).

\bibitem{boykin}
P. Oscar Boykin, T. Mor, V. Roychowdhury, F. Vatan, R. Vrijen, PNAS {\bf 99},
3388 (2002).

\bibitem{boehm_1}
H.P. Boehm, Angew. Chem. Int. Ed. {\bf 49}, 9332 (2010). 

\bibitem{boehm_2}
H.P. Boehm, R. Setton, and E. Stumpp, carbon {\bf 24}, 241 (1986).

\bibitem{graphene_book}
H. Aoki and M.S. Dresselhaus (Eds.), {\it Physics of Graphene}
(Springer, New York 2014).

\bibitem{graphene_review}
D.S.L. Abergel, V. Apalkov, J. Berashevich, K. Ziegler, and
T. Chakraborty, Adv. Phys. {\bf 59}, 261 (2010).

\bibitem{ssc}
T. Chakraborty and V.M. Apalkov, Solid State Commun. {\bf 175 - 176}, 
123 (2013).

\bibitem{xuefeng}
X.F. Wang and T. Chakraborty, Phys. Rev. B {\bf 81}, 081402(R)
(2010). 

\bibitem{abergel}
D.S.L. Abergel and T. Chakraborty, Nanotechnology {\bf 22}, 015203 (2011).

\bibitem{julia}
J. Berashevich and T. Chakraborty, Nanotechnology {\bf 21}, 355201 (2010).

\bibitem{abergel_ring}
D.S.L. Abergel, V.M. Apalkov and T. Chakraborty, Phys. Rev. B {\bf 78},
193405 (2008).

\bibitem{Recher76}
P. Recher, B. Trauzettel, A. Rycerz, Ya.M. Blanter, C.W.J. Beenakker,
and A.F. Morpurgo, Phys. Rev. B \textbf{76}, 235404 (2007).

\bibitem{russo}
S. Russo, J.B. Oostinga, D. Wehenkel, H.B. Heersche, S.S. Sobhani,
L.M.K. Vandersypen, and A.F. Morpurgo, Phys. Rev. B {\bf 77}, 085413 (2008).

\bibitem{gmitra}
M. Gmitra, S. Konschuh, C. Ertler, C. Ambrosch-Draxl, and J. Fabian,
Phys. Rev. {\bf 80}, 235431 (2009).

\bibitem{marchenko}
D. Marchenko, A. Varykhalov, M.R. Scholz, G. Bihlmayer, E.I. Rashba,
A. Rybkin, A.M. Shikin, and O. Rader, Nat. Commun. {\bf 3}, 1232 (2012).

\bibitem{rossi}
J. Zhang, C. Triola, and E. Rossi, Phys. Rev. Lett. {\bf 112}, 096802 (2014).

\end{thebibliography}
\end{document}